# Robust site-resolved quantum gates in an optical lattice via inhomogeneous control.

J. H. Lee[1], E. Montano[1], I. H. Deutsch[2], and P. S. Jessen[1]

[1]*Center for Quantum Information and Control, College of Optical Sciences and Department of Physics, University of Arizona, Tucson, Arizona 85721, USA.*
[2]*Center for Quantum Information and Control, Department of Physics and Astronomy, University of New Mexico, Albuquerque, New Mexico 87131, USA*

**Ultracold atoms in optical lattices[1] are an important platform for quantum information science, lending itself naturally to quantum simulation of many-body physics[2] and providing a possible path towards a scalable quantum computer[3,4]. To realize its full potential, atoms at individual lattice sites must be accessible to quantum control and measurement. This challenge has so far been met with a combination of high-resolution microscopes and resonance addressing that have enabled both site-resolved imaging[5,6,7] and spin-flips[8]. Here we show that methods borrowed from the field of inhomogeneous control[9,10] can greatly increase the performance of resonance addressing in optical lattices, allowing us to target arbitrary single-qubit gates on desired sites, with minimal crosstalk to neighboring sites and greatly improved robustness against uncertainty in the lattice position. We further demonstrate the simultaneous implementation of different gates at adjacent sites with a single global control waveform. Coherence is verified through two-pulse Ramsey interrogation, and randomized benchmarking[11] is used to measure an average gate fidelity of ~95%. Our control-based approach to reduce crosstalk and increase robustness is broadly applicable in optical lattices irrespective of geometry, and may be useful also on other platforms for quantum information processing, such as ion traps[12] and nitrogen-vacancy centers in diamond[13].**

Quantum simulations with the atom-lattice system generally explore many-body physics of condensed matter systems described by simple model Hamiltonians, e. g., the families of Hubbard[14] and Ising[15] models, and similar physics is typically relevant when exploring optical-lattice based architectures for universal quantum computing[16,17]. While optical lattices can have spatial periods from one-quarter to many times the optical wavelength, the need for site-to-site tunneling in such experiments tends to limit the workable lattice period to <1$\mu$m. Atoms in two-dimensional (2D) lattices of this type have been imaged using high numerical aperture optics with resolution close to the lattice spacing[6,7]. Other achievements include magnetic resonance imaging[18], optical microscopy in conjunction with numerical deconvolution[19], and imaging in a 3D lattice with ~5$\mu$m period[5]. Coherent quantum control of atoms at targeted sites requires even higher resolution than imaging if adjacent atoms are to remain unperturbed. Subwavelength resolution can be achieved with a tightly focused optical field that shifts the transition frequency of an atomic qubit relative to its neighbors, in combination with a frequency selective microwave pulse that implements the desired rotation of the targeted qubit[20]. In practice, such a resonance addressing scheme involves a tradeoff – for a sharply focused addressing field the frequency shift becomes overly sensitive to its alignment relative to the target site, while a softer focus leads to unwanted perturbations at adjacent sites. Faced with these difficulties, experiments have so far demonstrated only adiabatic spin flips[8], which are robust to small variations in the frequency shift but cannot manipulate coherence between the spin-up and spin-down states in the manner of universal quantum gates. Misalignment between the trapping and addressing fields can be avoided by incorporating a spatially varying qubit frequency shift into the lattice itself. The approach has been used to target quantum gates on subensembles of qubits located on one or the other side of the barrier in a lattice consiting of double wells, but does not lend itself to more general forms of addressing[21].

In this letter we explore the use of advanced quantum control to dramatically improve and extend the capabilities of resonance addressing. It is known from the theory of inhomogeneous control in Nuclear Magnetic Resonance that composite pulses can be designed to achieve any desired qubit response as function of one or more parameters[9,10,22]. Applying this idea to resonance addressing, we first map position onto frequency and then design a single phase-modulated microwave pulse to implement a unitary transformation that is a chosen function of frequency. An example might be a pulse that executes a quantum gate across a desired frequency and spatial interval, while at the same time doing nothing (the identity transformation) outside it. Such a "top-hat" response can greatly improve the robustness against misalignment between trapping and addressing fields, while at the same time suppressing perturbation of the adjacent sites.

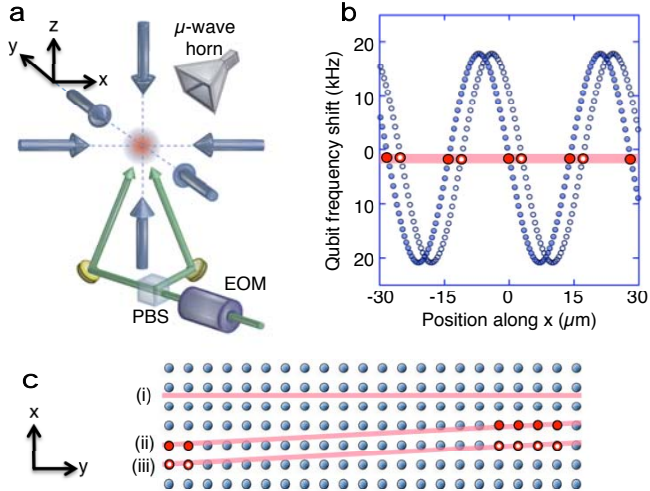

**Figure 1 | Trapping and addressing in an optical lattice.**
**a,** Atoms are trapped in a 3D lattice (blue), in the presence of a 1D addressing lattice (green) that can be translated using an electro-optic modulator (EOM) and polarization beamsplitter (PBS). **b,** Variation in the qubit transition frequency for sites in the trap lattice (blue dots). A microwave pulse of appropriate frequency (red line) addresses atoms at resonant sites (red dots). Translation of the addressing lattice (blue circles) targets different atoms (red circles). **c,** Schematic of trap lattice sites in the *x-y* plane. (i) When the trap and addressing lattices are perfectly aligned, the *y-z* plane addressed by the microwave frequency (red line) may not overlap with any sites. (ii) A small misalignment ensures there are resonant sites (red dots) for any microwave frequency and position of the addressing lattice. (iii) A translation of the addressing lattice by one trap lattice period still allows addressing of adjacent sites (red circles) along *x*.

To test the basic idea in the laboratory, we have implemented site-resolved resonance addressing along one dimension of a 3D optical lattice with a $\Lambda = 426$nm period, using a superimposed optical standing wave – the "addressing lattice" – as illustrated in Fig. 1a. The trap lattice is loaded with cesium atoms at a density of one atom per ~100 sites, each representing a qubit encoded in two spin states, $|\uparrow\rangle$ and $|\downarrow\rangle$. The addressing lattice is formed by two laser beams intersecting at a shallow angle, producing a frequency shift that varies sinusoidally along the *x* axis of the trap lattice as shown in Fig. 1b. Because the frequency shift is uniform in the *y-z* plane, this geometry addresses planes of ~$10^3$ atoms simultaneously, rather than single sites in the 3D trap lattice.

The addressing lattice is translated with nanometer accuracy along *x* by changing the relative optical phases between the addressing beams. Passive stability keeps jitter in the relative position of the trap and addressing lattices below 15nm on a timescale of several seconds, sufficient that it can be regarded as constant during a single run of the experiment, while on timescales from minutes to hours the lattice positions may drift by several microns. Note that if the planes of the trap and addressing lattices are parallel, this slow drift implies that many experimental runs will have no trap sites within the frequency interval targeted by a given microwave pulse. (Fig. 1c-i). While one could in principle stabilize and actively control the relative position[8], it is more convenient for our purpose if each run of the experiment samples the full range of relative positions and qubit transition frequencies that our pulse is designed to address. This can be achieved by tilting the planes of the addressing lattice slightly relative to those of the trap lattice, which assures there will always be subsets of trap sites that fall within as well as outside the targeted interval, regardless of drift in the relative positions of the trap and addressing lattices (Fig. 1c-ii). When combined with a suitable preselection protocol as described below, this is precisely the situation needed to observe the inhomogeneous response of our composite pulses and evaluate the performance of the corresponding quantum gates.

To prepare a well defined starting point, we preselect a subset of atoms by flipping their spins from $|\uparrow\rangle$ to $|\downarrow\rangle$ with a resonant microwave pulse and removing the remaining atoms in $|\uparrow\rangle$. The result is an ensemble whose distribution of resonance frequencies and positions along *x* reflects the spin-flip probability in a pulse with a Gaussian frequency spectrum. We can construct a resonance image of this distribution over many repetitions of the experiment, by following the "preparation" pulse with a variable translation of the addressing lattice, a second "imaging" pulse, and a measurement of the number of atoms returned to the $|\uparrow\rangle$ state. Figure 2A shows such an image, together with a Gaussian fit with standard deviation $\sigma = 80$nm. The image width is given by the convolution of the (identical) preparation and imaging pulses, and indicates a spatial resolution of ~56nm, well below the lattice period. It is also possible to prepare atoms at adjacent sites along *x* by applying a sequence of preparation pulses with identical frequency and separated by appropriate translations of the addressing lattice (Fig. 1c-iii). For example, a resonance image of atoms at three adjacent sites is shown in Fig. 2C-a.

More advanced quantum control can be performed with phase modulated microwave pulses. These composite pulses consist of a train of *N* square pulses with variable phases $\{\varphi_j\}$ that have been computer optimized so the overall transformation accomplishes a desired objective[9,10]. This can be either a spin rotation on a fixed input state, $|\downarrow\rangle \rightarrow \alpha(\delta)|\downarrow\rangle + \beta(\delta)|\uparrow\rangle$, or a full unitary transformation (quantum gate) $W(\delta)$ that varies in a prescribed way with the frequency shift $\delta$ of the qubit resonance relative to the microwave frequency. Figure 2B shows the response to a composite pulse designed to flip spins uniformly across a targeted frequency region and leave them unaffected elsewhere. We can test the performance of a similar top-hat pulse whose three regions have been optimized to overlap with the three-site atom distribution in Fig. 2C-a. Figure 2C-b shows the distribution after the pulse, indicating a complete, robust spin flip at the central site, and a complete, robust absence of spin flips at the adjacent sites. For comparison, Fig. 2C-c shows the result of applying a single Gaussian pulse (identical to the preparation pulses) resonant at the center of the distribution. In this case some spins at the central site are unaffected, while some spins at adjacent sites have been flipped. Adjusting the width of the pulse allows a tradeoff between the two types of error, but performance never approaches Fig. 2C-b.

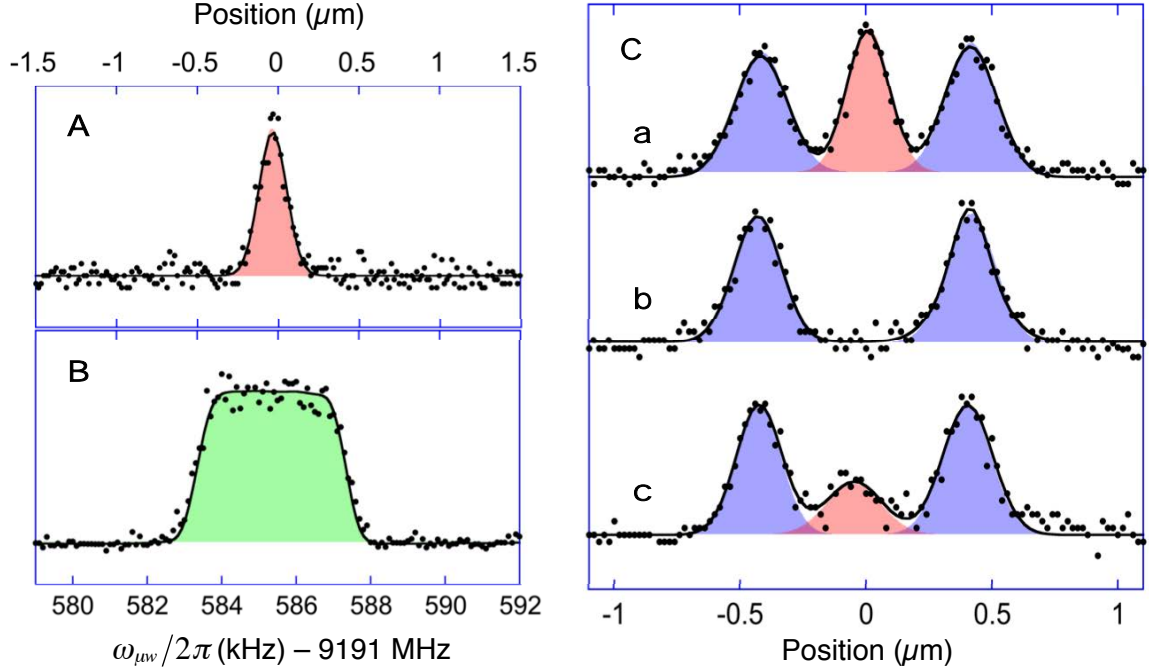

**Figure 2 | Resonance imaging and addressing of single planes. A**, Resonance image showing a distribution of atoms prepared at in a single plane as indicated in fig 1c-ii. **B,** Spin flip probability for a composite microwave pulse with a "top hat" response as function of frequency. The frequency and position axes are matched between A) and B). **C**, Multi-site imaging and addressing. **a**, Preparation of atoms in three adjacent planes. **b**, Selective spin flip of atoms at the central site, performed using a composite pulse with a top hat response similar to, but narrower than B) **c**, Imperfect spin flip of atoms at the central site, performed using a Gaussian pulse. Solid dots represent data and lines represent fits. In A) and C) the shaded areas show the separate Gaussians fits associated with atoms at each site.

The performance of unitary quantum gates can be evaluated with a two-pulse version of the approach above. We begin by applying a composite pulse that implements a $\pi/2$ rotation around the **i**-axis of the Bloch sphere in the central frequency region and the identity elsewhere. Shifting the overall phase of the pulse by $\phi$ changes the axis of rotation to $\cos(\phi)\mathbf{i}+\sin(\phi)\mathbf{j}$, and two pulses in sequence lead to a spin-flip probability $\sim \cos^2(\phi/2)$, similar to the $\phi$-dependent interference in a two-path interferometer. Figure 3a-d shows a sequence of resonance images and corresponding populations remaining in $|\downarrow\rangle$ as function of the phase $\phi$. The clear interference for atoms in the central region and the lack of $\phi$-dependence for atoms in the adjacent regions demonstrates that the pulse functions as intended and coherence is preserved.

A second example of a two-pulse experiment is shown in Fig. 3e-h, in this case for frequency regions that correspond to three adjacent sites as in Fig. 2C. Here the first pulse implements a Hadamard gate (rotation by $\pi$ around the $(\mathbf{i}+\mathbf{k})/\sqrt{2}$ axis) on the central site and $\pi/2$ rotations on the adjacent sites. The second pulse implements identical $\pi/2$ rotations at all three sites, leading to interference patterns similar to Fig. 3c but with a 90° shift at the central site. This data set shows explicitly that the gate operations are coherent on all three sites, and also demonstrates the freedom to perform independent gates simultaneously at adjacent sites with a single composite pulse. The contrast of the interference patterns in Figs. 3f-h provides information about the fidelity of the various transformations. Assuming that gate errors are uncorrelated and independent of $\phi$, we can estimate the fidelity of one gate in a pair as $\mathcal{F} = \sqrt{S_{max}/(S_{min}+S_{max})}$, where $S_{min}$ and $S_{max}$ are the minimum and maximum values of the interference signal. For the data in Fig. 3f-h this yields an average gate fidelity of $\mathcal{F} = 0.96$.

A more comprehensive test of gate fidelity is best performed by randomized benchmarking[11]. For the 4ms pulses used in Fig. 3, the total duration of the necessary pulse sequences exceeds the time available in our experiment, but useful information can still be obtained by benchmarking a set of pulses that have been rescaled and shortened to 1ms duration (see Methods). Figure 4 shows the fidelity for sequences of rescaled pulses that were originally designed to implement computational gates (CGs) on a target lattice site and the identity at neighboring sites.

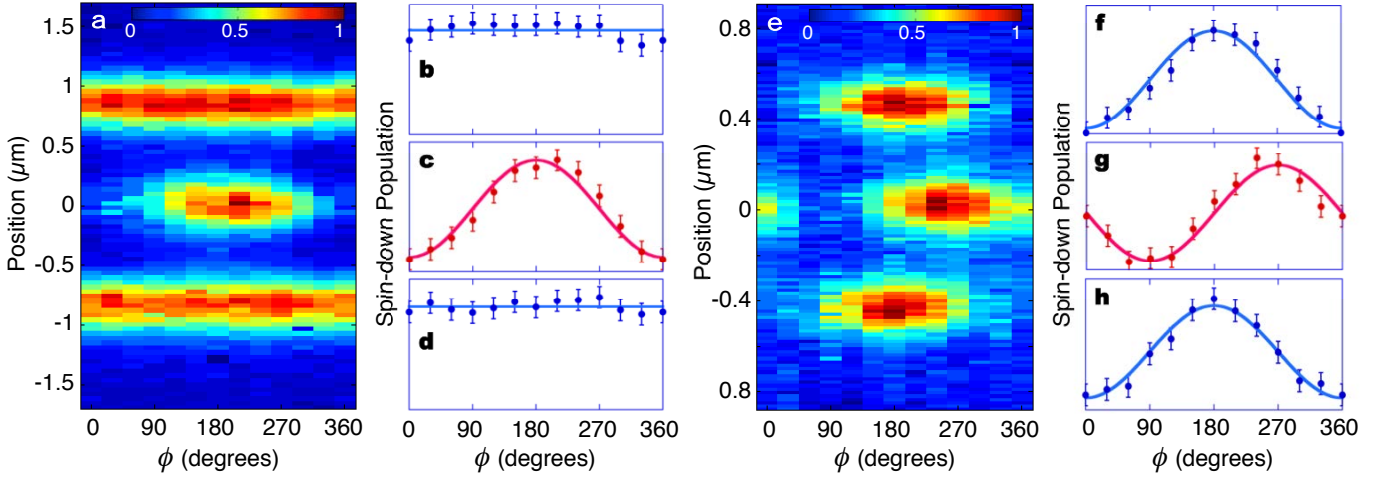

**Figure 3 | Unitary quantum gates. a**, Sequence of color coded resonance images (similar to Fig. 2C), showing the coherent action of a pair of composite pulses. Each pulse implements a $\pi/2$ rotation on the central atom distribution, with full spin flips occurring at $\phi = 0°$, 360°, and the identity at $\phi = 180°$. Both the individual pulses and the pair always implement the identity on the adjacent atom distributions. **b-d**, Population remaining in the initial $|\downarrow\rangle$ state as function of relative pulse phase, for each distribution. The populations (dots) are determined from the areas of Gaussian fits as in Fig. 2C. Solid lines are fits to the interference pattern, and error bars are estimated as one standard deviation of the residuals from the fit. **e-h**, Sequence of resonance images and interference patterns similar to **a-d**), but with atom distributions chosen to coincide with adjacent sites along $x$. The first composite pulse here implements a Hadamard pulse at the central site and a $\pi/2$ rotation at the adjacent sites, the second pulse implements $\pi/2$ rotations at all three sites.

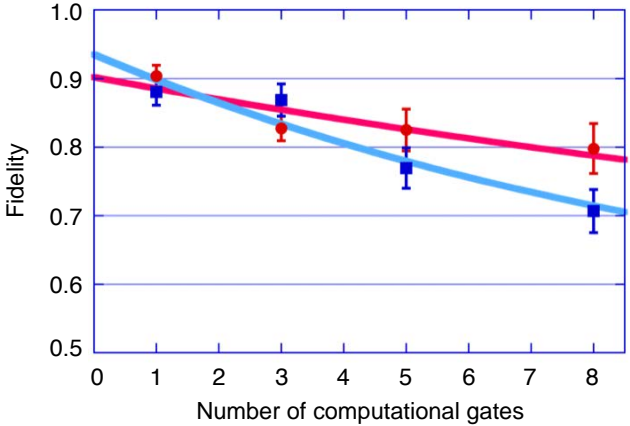

**Figure 4 | Randomized benchmarking.** Overall fidelities achieved by sequences of composite pulses that apply computational and randomizing gates to the central atom distribution (red), and repeated identities to adjacent ones (blue). Dots and squares are experimental data and lines are fits. Error bars are one standard deviation, estimated from the spread of fidelities seen in different sequences.

Fits yield an average fidelity $F_{CG} = 0.98$ for the central region, and a fidelity per identity of $F_I = 0.96$ for the adjacent regions. In the absence of external perturbations, e.g., magnetic fields and light scattering, one would expect similar fidelities for the site resolved gates implemented by the original pulse. On the other hand, if gate errors were entirely due to external factors, one might expect four times the error for the original pulses. This implies a fidelity in the range $F_{CG} = 0.92 - 0.98$ for the site resolved quantum gates in Fig. 3. This is consistent with the $F \sim 0.96$ obtained from the two-pulse interference patterns, especially since the latter estimate includes a non-negligible contribution from initialization and readout errors.

Our laboratory realization of robust, site-resolved quantum gates through quantum control points to further experiments aimed at control of atoms in optical lattices. In our geometry, increasing the gradient of the frequency shift will allow the implementation of faster quantum gates with shorter pulses, while increasing the pulse amplitude and number of phase steps will allow the simultaneous implementation of independent gates across a much larger number of lattice sites and reduce the overhead associated with serial resonance addressing. Most importantly, the use of a focused addressing field[8] should make it straightforward to address a single site in a 2D or 3D optical lattice. From a control perspective this task is simpler than the one undertaken here, since one only has to consider the atomic response in two regions – one covering frequency shifts near the focus, and one covering the remaining range of frequency shifts down to zero. Further developments might use site resolved single atom control to "activate"[21] qubits for detection, or for the implementation of localized two-qubit gates though cold collisions[16,17] or dipole-dipole interactions of Rydberg atoms[23,24]. Ultimately the ability to simultaneously execute quantum logic gates on different qubits could substantially reduce the computational time-complexity of quantum algorithms. With such parallelization, the circuits for the Quantum Fourier Transform and encoding/decoding of quantum-error correcting codes with $O(n)$ qubits can be compressed to $O(\log n)$ operations[25,26]. Other paradigms such as measurement-based quantum computation also benefit form parallelization; almost all gates can be implemented simultaneously in a single control and measurement step[27].

**METHODS SUMMARY**
Our 3D trap lattice consists of three orthogonal, linearly polarized standing waves detuned 140 GHz from resonance. We load the trap lattice with ~$10^7$ atoms from a magneto-optic trap and optical molasses, and use sideband cooling[28,29] to prepare them in the $|\uparrow\rangle$ state with mean vibrational excitation $\bar{n} \approx 0.01$ along $x$. A bias magnetic field isolates the qubit transition frequency from others in the hyperfine ground state, and the bias and background magnetic fields are stabilized so the qubit frequency uncertainty is <50Hz[30]. Populations of the qubit states are measured via Stern-Gerlach analysis. The addressing lattice is formed by two plane waves with orthogonal linear polarizations, intersecting at an angle of 1.74° to produce a light shift with a period of 28 $\mu$m, or 65.9 times that of the trap lattice. Our microwave field is radiated by a pair of horn antennae adjusted to minimize field inhomogeneity across the atomic ensemble. The pulses used for preparation and imaging are Gaussian in the time and frequency domains, with rms widths of 0.5ms and 225Hz. Robust spin-flips and quantum gates are implemented with composite pulses. These consist of a train of square pulses whose individual phases are computer optimized to minimize the distance between the desired and actual transformations across a targeted bandwidth. These pulses can be rescaled by increasing the amplitude and shortening the duration by a factor $\kappa$, thereby changing the frequency dependence of the corresponding quantum gate, $W(\delta) \to W(\delta/\kappa)$, without otherwise affecting the control fidelity. Randomized benchmarking[11] is performed by applying sequences of alternating $\pi/2$ "computational" gates (CG's) and "Pauli randomizing" gates (PG's), and measuring the decay in the overall fidelity as a function of the number of computational gates

**Acknowledgements** This work was supported by the US National Science Foundation Grants PHY-0903692, 0903930, 0903953, 0969371.

**Author Contributions** JHL and EM performed the experiment and data analysis. IHD and PSJ provided ideas and advice, and PSJ supervised the work.

**Author Information** Reprints and permissions information is available at www.nature.com/reprints. The authors declare no competing financial interests. Correspondence and requests for materials should be addressed to P. S. J. (poul.jessen@optics.arizona.edu).

## METHODS

**Atom trapping, preparation, control and measurement.**
Our trap lattice is formed by three pairs of counter propagating laser beams with parallel linear polarizations, tuned 140GHz above the $6S_{1/2}(f=4) \rightarrow 6P_{3/2}(f'=5)$ hyperfine transition at 852nm. The depth of the optical potential is ~40 $\mu$K, corresponding to a trap vibrational frequency of 20kHz. We load the trap lattice with $10^7$ atoms from a magneto-optic trap and optical molasses, with a density of one atom per 100 sites in a volume with a diameter of ~35 addressing lattice periods. Qubits are encoded in states $|\downarrow\rangle = |f=3, m_f=-3\rangle$ and $|\uparrow\rangle = |f=4, m_f=-4\rangle$ in the hyperfine ground state. We use sideband cooling[28,29] to prepare atoms in the $|\uparrow\rangle$ state, with mean vibrational excitation $\bar{n} \approx 0.01$ along $x$ (excitation along $y$ and $z$ is unimportant). A bias magnetic field separates the qubit transition frequency from others in the ground manifold, and the combined bias, DC and AC background magnetic fields are stabilized to better than $20\mu$G using an approach similar to ref. 30. The resulting frequency variation in the qubit transition frequency is ~50H$z$. The atomic qubits are driven with 9.2GHz microwaves from two horn antennae adjusted to improve the field homogeneity across the atom cloud. Selective removal of atoms in the $|\uparrow\rangle$ state is accomplished with radiation pressure from a resonant laser beam. Populations of the $|\uparrow\rangle$ and $|\downarrow\rangle$ states are obtained via Stern-Gerlach measurement.

**Addressing lattice**
The addressing lattice is formed by two plane waves with orthogonal linear polarizations, intersecting at an angle of 1.74°. The light shift in this configuration is equivalent to a fictitious magnetic field, with the steepest gradient where its value is near zero (Fig. 1), the most favorable situation for resonance imaging and addressing. The addressing lattice period is $28\mu$m, corresponding to 65.9 periods of the trapping lattice. For perfectly aligned trap and addressing lattices, such incommensurate periods would slightly shift the relative position of the trap sites from period to period of the addressing lattice, but this becomes irrelevant for misaligned lattices as discussed in the main text. Accurate calibration of the frequency-to-position relationship in the addressing lattice is performed by moving atoms in the trap lattice an integer number of sites through polarization rotation[3,4,16] (see Supplemental Materials).

## Microwave pulse design

Resonance preparation and imaging is performed with microwave pulses having fixed frequency and phase, and a Gaussian envelope with 0.5ms rms width in the time domain. The corresponding power spectrum is also Gaussian with an rms width of 225Hz. More advanced control is performed with composite microwave pulses consisting of a train of $N$ square pulses having common amplitude $A$, duration $T$ and frequency $\omega_{\mu w}$, but with phases $\varphi_j$ that vary between each pulse in the train. We keep $\omega_{\mu w}$ fixed and use the set of phases $\{\varphi_j\}$, along with $A$, $T$, and the detuning $\delta = \omega_{qubit} - \omega_{\mu w}$ from qubit resonance, as our control parameters. In that case the composite rotation implemented by the pulse train corresponds to an overall unitary transformation $U_N(\{\varphi_j\}, A, T, \delta)$. We can then use standard numerical techniques to search for control parameters that achieve a desired objective, e.g., implementing a quantum logic gate $W(\delta)$ that is a prescribed function of $\delta$. This is done by defining a cost function, in our case the distance between the target and actual unitary matrices, averaged over a frequency band $\Delta$,

$$C(\{\varphi_j\}, A, T) = \frac{1}{\Delta} \int_{-\Delta/2}^{\Delta/2} d\delta \|W(\delta) - U_N(\{\varphi_j\}, A, T, \delta)\|,$$

where $\|W - U\| = \sqrt{(W-U)^\dagger(W-U)}$ is the Hilbert-Schmidt distance between $W$ and $U$. Minimizing $C$ then finds a set of values $\{\varphi_j\}, A, T$ such that $U_N(\{\varphi_j\}, A, T, \delta) \approx W(\delta)$. Note that this cost function includes the overall phase between $U$ and $W$; since this phase is not physically meaningful the control task could be simplified by instead maximizing the frequency average of $|Tr(U^\dagger W)|$. Different control objectives can be achieved by substituting an appropriate cost function, e. g. for frequency dependent spin flips the cost function is the infidelity averaged over the relevant frequency interval, $C(\{\varphi_j\}, A, T) = 1 - |\langle \psi(\delta) | U_N(\{\varphi_j\}, A, T, \delta) | \downarrow \rangle|^2$, where $|\psi(\delta)\rangle = \alpha(\delta)|\downarrow\rangle + \beta(\delta)|\uparrow\rangle$ is the desired final spin state as function of $\delta$. See Supplemental Materials for additional details.

## Pulse rescaling

A square pulse with Rabi frequency $\Omega_0 \propto A$, duration $T$, detuning $\delta$ and phase $\varphi_j$ will rotate the qubit Bloch vector by an angle $\theta = \Omega T$ around an axis $\mathbf{q} = (\Omega_0/\Omega)\cos(\varphi)\mathbf{i} + (\Omega_0/\Omega)\sin(\varphi)\mathbf{j} + (\delta/\Omega)\mathbf{k}$, where $\Omega = \sqrt{\Omega_0^2 + \delta^2}$. It follows that replacing $\Omega_0 \to \kappa\Omega_0$, $T \to T/\kappa$, and $\delta \to \kappa\delta$ leaves the angle and axis of rotation unchanged. Extending this to an entire pulse train, we see that $U_N(\{\varphi_j\}, A, T, \delta) = U_N(\{\varphi_j\}, \kappa A, T/\kappa, \kappa\delta)$. This implies that if a pulse train implements $U_N(\{\varphi_j\}, A, T, \delta) \approx W(\delta)$, then a rescaled pulse train implements $U_N(\{\varphi_j\}, \kappa A, T/\kappa, \delta) \approx W(\delta/\kappa)$, with the same value of the cost function when averaged over a bandwidth $\kappa\Delta$. As a result, a composite pulse can be rescaled to stretch or compress it in the time domain while compressing or stretching it in the frequency domain, without otherwise changing the fidelity with which it implements the desired objective (the control fidelity).

## Randomized benchmarking

Following ref. 11, we perform randomized benchmarking by initializing atoms in the $|\downarrow\rangle$ state, applying $l$ successive pairs of $\pi/2$ "computational" gates (CG's) and "Pauli randomizing" gates (PG's), and reading out the overall fidelity with which the qubit is returned to the $|\downarrow\rangle$ state. The sequence is repeated with random choices of CG's and PG's to obtain average overall fidelities as function of $l$. This data is then fitted with a function $F = [1 + (1-\varepsilon_0)(1-2\varepsilon)^l]/2$, where $\varepsilon_0$ is the combined initialization and readout error and $\varepsilon$ is the average error per computational gate. To fit a sufficient number of composite pulses into the available time window we shorten them to 1ms (rescaling by $\kappa = 4$) Provided that errors are dominated by imperfections in the control fields, and that the pulses are tested on a broadened version of the distribution in Fig. 2C, benchmarking will then provide a valid measure of the experimental gate fidelity.

## SUPPLEMENTARY DISCUSSION

**Translations and frequency shifts of the addressing lattice.** Our addressing lattice is translated along the *x* axis by shifting the relative phase between its plane wave components; this phase shift is produced by applying a voltage to an Electro-Optic Modulator inserted in the beam path as shown in Fig. 1a. To prepare atoms at adjacent sites in the manner illustrated in Fig. 1c, it is essential to have an accurate calibration of translation in units of trap lattice periods as function of the EOM control voltage. This can be accomplished by preparing a sample of atoms in a single *y-z* plane at position $x_0$, displacing them by a known distance $\Delta x$ along *x*, constructing a resonance image as described in the main text, and noting the EOM voltage at the center of the image where the addressing lattice has been translated by the same distance $\Delta x$ along *x*. As illustrated in Fig. S1a, we can move our atoms by changing the angle $\theta$ between the linear polarizations of the beams that form the trap lattice along *x*, with angles $\theta = \{-360°, -180°, 0°, 180°, 360°\}$ corresponding to atom displacements $\Delta x = \{-\Lambda, -\Lambda/2, 0, \Lambda/2, \Lambda\}$, where $\Lambda = 426$nm is the trap lattice period [31]. Figure S1b shows resonance images obtained for various polarization angles $\theta$. From the EOM voltages at the image centers (Fig. S1c) we infer that the addressing lattice is translated by one trap lattice period per 164V. This is consistent with the value expected from independent measurements of the EOM phase shift versus voltage and the intersection angle of 1.74° between the addressing lattice beams.

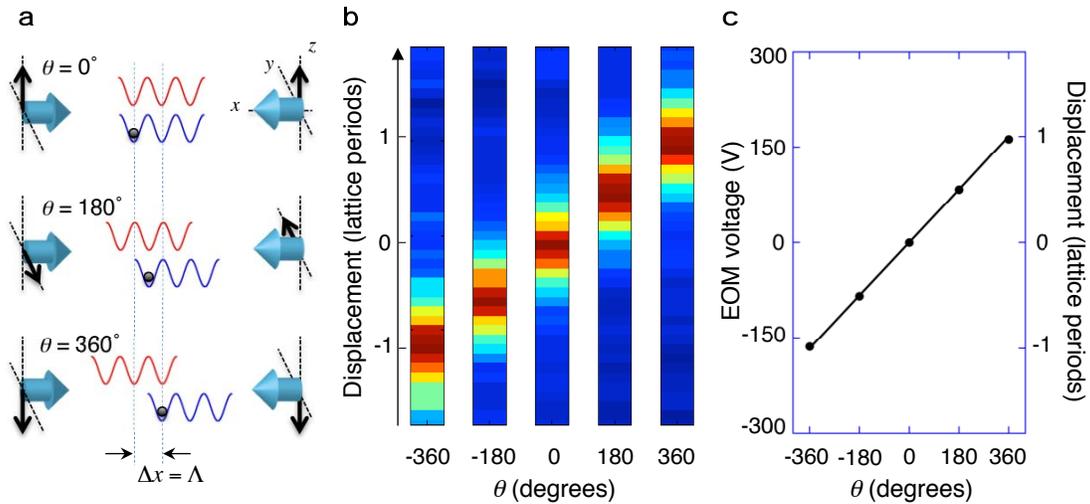

**Figure S1 | Calibrating translations of the addressing lattice. a**, Atoms are displaced by a known distance along the *x* axis by rotating the polarizations (black arrows) of the trap lattice beams (blue arrows). This shifts the potentials for the $|\uparrow\rangle$ (red) and $|\downarrow\rangle$ (blue) qubit states in opposite directions. **b**, Resonance images obtained for atoms displaced by multiples of $\Lambda/2$. **c**, Measured relationship between the angle $\theta$ between the trap lattice polarizations, the atom displacement $\Delta x$, and the EOM voltage corresponding to the center of the resonance image.

Once we have calibrated the addressing lattice displacement versus EOM voltage, the spatial gradient of the frequency shift produced by the addressing lattice can be determined. To do so, we first prepare atoms in a single plane near the point of steepest gradient, apply a known Zeeman shift $\Delta\omega$ of the qubit transition frequency with an external magnetic field (Fig. S2A), and construct a resonance image (Fig. S2B). Note that our ensemble will contain atoms at points where the gradient is positive as well as negative, and that a frequency shift $\Delta\omega \neq 0$ will therefore result in a double-peaked resonance image (Fig. S2B-b). Recording the separation between these peaks versus the Zeeman shift provides an accurate measurement of the spatial gradient of the frequency shift in the addressing lattice (Fig. S2C).

**Microwave pulses**
We use an inexpensive microwave source as depicted in Fig. S3a, combining an HP8672A Synthesized Signal Generator that supplies a carrier at 9.2GHz, and a Tabor WW2572A-2 Direct Digital Synthesis Arbitrary Waveform Generator that supplies a signal at 30MHz. The two are mixed in a single-sideband mixer whose output is preamplified, split in two, passed through a pair of 2W power amplifiers, and radiated by a pair of horn antennae with 15dB gain. A low-power microwave switch between the HP8672A and the mixer is used for digital on/off control, while amplitude, frequency and phase modulation of the WW2572A-2 under internal arbitrary waveform control is used to correspondingly modulate the sideband used to drive the atomic qubits. The overall system delivers microwave fields with amplitude sufficient to drive the qubit transition with Rabi frequencies up to 40kHz. The amplitude modulation capability is used when we generate preparation and imaging pulses with Gaussian envelope in the time domain, while the phase modulation capability is used when we generate composite pulses for more advanced inhomogeneous control. An example of the time-dependent qubit Rabi frequency (proportional to μw amplitude) and phase of one of these composite pulses is shown in Figs. S3b&c.

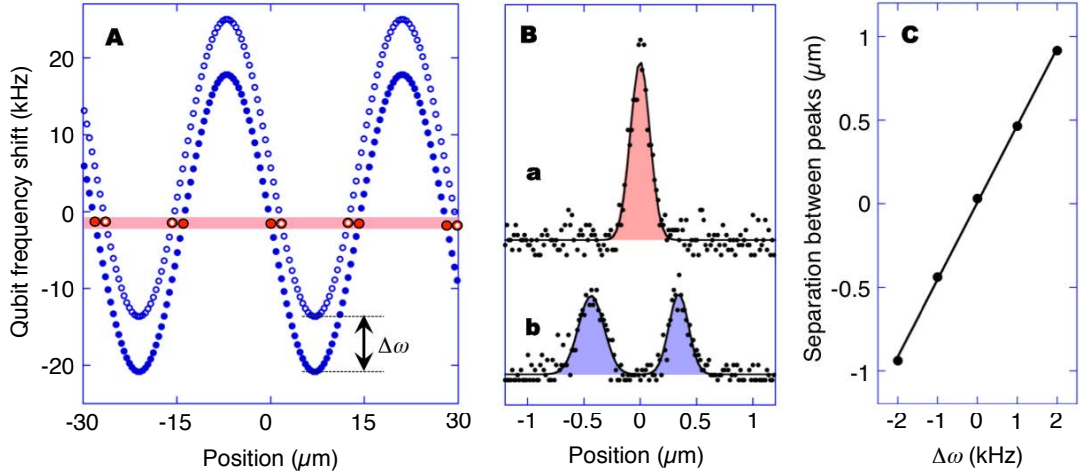

**Figure S2 | Spatial gradient of the addressing lattice. A**, The qubit transition frequencies at the trap lattice sites (blue dots) can be Zeeman shifted by an amount Δω with an external magnetic field (blue circles). A microwave pulse initializes atoms at the point of steepest gradient (red dots). After the Zeeman shift the imaging pulse is resonant at positions to the left and right of the atoms (red circles), and the addressing lattice must be translated to the right or left to bring them back into resonance. **B-a**, Resonance image of the initial atom distribution (Δω = 0). **B-b**, Double-peaked resonance image of the atom distribution after a Zeeman shift Δω ≠ 0. **C**, Separation between the resonance peaks in B-b as function of the Zeeman shift Δω. The orresponding gradient is 4.35kHz/μm.

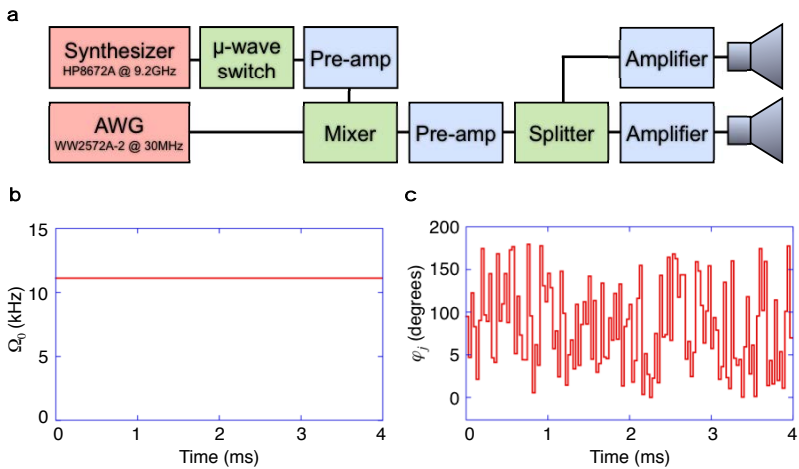

**Figure S3 | Microwave pulses. a**, Our inexpensive microwave source consists of a 9.2GHz Synthesizer and 30MHz Arbitrary Waveform Generator (AWG), mixed and amplified and radiated by a pair of horn antennae. **b**, Qubit Rabi frequency (proportional to microwave amplitude) versus time during our composite pulses. **c**, Example of the microwave phase modulation versus time during one of our composite pulses.

[31] Mandel, O., Greiner, M., Widera, A, Rom, T., Hänsch, T. W. & Bloch, I. Coherent transport of neutral atoms in spin-dependent optical lattice potentials. Phys. Rev. Lett. **91**, 010407 (2003).